\documentstyle[12pt]{article}
\advance \topmargin by -\headheight
\advance \topmargin by -\headsep

\evensidemargin \oddsidemargin
\begin{document}

\topmargin -.6in

\def\rf#1{(\ref{eq:#1})}
\def\lab#1{\label{eq:#1}}
\def\nonu{\nonumber}
\def\br{\begin{eqnarray}}
\def\er{\end{eqnarray}}
\def\be{\begin{equation}}
\def\ee{\end{equation}}
\def\eq{\!\!\!\! &=& \!\!\!\! }
\def\foot#1{\footnotemark\footnotetext{#1}}
\def\lb{\lbrack}
\def\rb{\rbrack}
\def\llangle{\left\langle}
\def\rrangle{\right\rangle}
\def\blangle{\Bigl\langle}
\def\brangle{\Bigr\rangle}
\def\llb{\left\lbrack}
\def\rrb{\right\rbrack}
\def\Blb{\Bigl\lbrack}
\def\Brb{\Bigr\rbrack}
\def\lcurl{\left\{}
\def\rcurl{\right\}}
\def\({\left(}
\def\){\right)}
\def\v{\vert}                     
\def\bv{\bigm\vert}               
\def\Bgv{\;\Bigg\vert}            
\def\bgv{\bigg\vert}              
\def\lskip{\vskip\baselineskip\vskip-\parskip\noindent}
\def\mskp{\par\vskip 0.3cm \par\noindent}
\def\sskp{\par\vskip 0.15cm \par\noindent}
\def\bc{\begin{center}}
\def\ec{\end{center}}
\def\Lbf#1{{\Large {\bf {#1}}}}
\def\lbf#1{{\large {\bf {#1}}}}
\relax

\def\tr{\mathop{\rm tr}}                  
\def\Tr{\mathop{\rm Tr}}                  
\newcommand\partder[2]{{{\partial {#1}}\over{\partial {#2}}}}
\newcommand\funcder[2]{{{\delta {#1}}\over{\delta {#2}}}}
\newcommand\Bil[2]{\Bigl\langle {#1} \Bigg\vert {#2} \Bigr\rangle}  
\newcommand\bil[2]{\left\langle {#1} \bigg\vert {#2} \right\rangle} 
\newcommand\me[2]{\langle {#1}\vert {#2} \rangle} 

\newcommand\sbr[2]{\left\lbrack\,{#1}\, ,\,{#2}\,\right\rbrack} 
\newcommand\Sbr[2]{\Bigl\lbrack\,{#1}\, ,\,{#2}\,\Bigr\rbrack} 
\newcommand\pbr[2]{\{\,{#1}\, ,\,{#2}\,\}}       
\newcommand\Pbr[2]{\Bigl\{ \,{#1}\, ,\,{#2}\,\Bigr\}}  
\newcommand\pbbr[2]{\lcurl\,{#1}\, ,\,{#2}\,\rcurl}  

\def\a{\alpha}
\def\b{\beta}
\def\c{\chi}
\def\d{\delta}
\def\D{\Delta}
\def\eps{\epsilon}
\def\vareps{\varepsilon}
\def\g{\gamma}
\def\G{\Gamma}
\def\grad{\nabla}
\def\h{{1\over 2}}
\def\l{\lambda}
\def\L{\Lambda}
\def\m{\mu}
\def\n{\nu}
\def\ov{\over}
\def\om{\omega}
\def\O{\Omega}
\def\p{\phi}
\def\P{\Phi}
\def\pa{\partial}
\def\tpa{{\tilde \partial}}
\def\pr{\prime}
\def\ra{\rightarrow}
\def\lra{\longrightarrow}
\def\s{\sigma}
\def\S{\Sigma}
\def\t{\tau}
\def\th{\theta}
\def\Th{\Theta}
\def\z{\zeta}
\def\ti{\tilde}
\def\wti{\widetilde}
\def\one{\hbox{{1}\kern-.25em\hbox{l}}}

\def\cA{{\cal A}}
\def\cB{{\cal B}}
\def\cC{{\cal C}}
\def\cD{{\cal D}}
\def\cE{{\cal E}}
\def\cH{{\cal H}}
\def\cL{{\cal L}}
\def\cM{{\cal M}}
\def\cN{{\cal N}}
\def\cP{{\cal P}}
\def\cQ{{\cal Q}}
\def\cR{{\cal R}}
\def\cS{{\cal S}}
\def\cU{{\cal U}}
\def\cV{{\cal V}}
\def\cW{{\cal W}}
\def\cY{{\cal Y}}

\def\phanta{\phantom{aaaaaaaaaaaaaaa}}
\def\phantb{\phantom{aaaaaaaaaaaaaaaaaaaaaaaaa}}
\def\phantc{\phantom{aaaaaaaaaaaaaaaaaaaaaaaaaaaaaaaaaaa}}

 \def\Winf{{\bf W_\infty}}               
\def\Win1{{\bf W_{1+\infty}}}           
\def\nWinf{{\bf {\hat W}_\infty}}       
\def\PsDA{\Psi{\cal DO}}
\def\Intres{\int dx\, {\rm Res} \; }

\def\KP3{{\bf KP_{2+1}}}
\def\mKP3{{\bf mKP_{2+1}}}
\def\KPm{{\bf (m)KP_{2+1}}}
\def\KPt{{\bf KP_{1+1}}}
\def\mKPt{{\bf mKP_{1+1}}}

\newcommand{\nit}{\noindent}
\newcommand{\ct}[1]{\cite{#1}}
\newcommand{\bi}[1]{\bibitem{#1}}
\newcommand\PRL[3]{{\sl Phys. Rev. Lett.} {\bf#1} (#2) #3}
\newcommand\NPB[3]{{\sl Nucl. Phys.} {\bf B#1} (#2) #3}
\newcommand\NPBFS[4]{{\sl Nucl. Phys.} {\bf B#2} [FS#1] (#3) #4}
\newcommand\CMP[3]{{\sl Commun. Math. Phys.} {\bf #1} (#2) #3}
\newcommand\PRD[3]{{\sl Phys. Rev.} {\bf D#1} (#2) #3}
\newcommand\PLA[3]{{\sl Phys. Lett.} {\bf #1A} (#2) #3}
\newcommand\PLB[3]{{\sl Phys. Lett.} {\bf #1B} (#2) #3}
\newcommand\JMP[3]{{\sl J. Math. Phys.} {\bf #1} (#2) #3}
\newcommand\PTP[3]{{\sl Prog. Theor. Phys.} {\bf #1} (#2) #3}
\newcommand\SPTP[3]{{\sl Suppl. Prog. Theor. Phys.} {\bf #1} (#2) #3}
\newcommand\AoP[3]{{\sl Ann. of Phys.} {\bf #1} (#2) #3}
\newcommand\RMP[3]{{\sl Rev. Mod. Phys.} {\bf #1} (#2) #3}
\newcommand\PR[3]{{\sl Phys. Reports} {\bf #1} (#2) #3}
\newcommand\FAP[3]{{\sl Funkt. Anal. Prilozheniya} {\bf #1} (#2) #3}
\newcommand\FAaIA[3]{{\sl Functional Analysis and Its Application} {\bf #1}
(#2) #3}
\def\InvM#1#2#3{{\sl Invent. Math.} {\bf #1} (#2) #3}
\newcommand\LMP[3]{{\sl Letters in Math. Phys.} {\bf #1} (#2) #3}
\newcommand\IJMPA[3]{{\sl Int. J. Mod. Phys.} {\bf A#1} (#2) #3}
\newcommand\TMP[3]{{\sl Theor. Mat. Phys.} {\bf #1} (#2) #3}
\newcommand\JPA[3]{{\sl J. Physics} {\bf A#1} (#2) #3}
\newcommand\JSM[3]{{\sl J. Soviet Math.} {\bf #1} (#2) #3}
\newcommand\MPLA[3]{{\sl Mod. Phys. Lett.} {\bf A#1} (#2) #3}
\newcommand\JETP[3]{{\sl Sov. Phys. JETP} {\bf #1} (#2) #3}
\newcommand\JETPL[3]{{\sl  Sov. Phys. JETP Lett.} {\bf #1} (#2) #3}
\newcommand\PHSA[3]{{\sl Physica} {\bf A#1} (#2) #3}
\newcommand\PHSD[3]{{\sl Physica} {\bf D#1} (#2) #3}

\newcommand\hepth[1]{{\sl hep-th/#1}}

\begin{titlepage}
\vspace*{-1cm}
\noindent
June, 1994 \hfill{BGU-94 / 15 / June- PH}\\
\phantom{bla}
\hfill{UICHEP-TH/94-6}\\
\phantom{bla}
\hfill{hep-th/9407017}
\\
\begin{center}
{\large {\bf Two-Matrix String Model as\\
Constrained (2+1)-Dimensional Integrable System}}
\end{center}
\vskip .3in
\begin{center}
{ H. Aratyn\footnotemark
\footnotetext{Work supported in part by the U.S. Department of Energy
under contract DE-FG02-84ER40173}}
\par \vskip .1in \noindent
Department of Physics \\
University of Illinois at Chicago\\
845 W. Taylor St.\\
Chicago, IL 60607-7059, {\em e-mail}:
aratyn@uic.edu \\
\par \vskip .3in
{ E. Nissimov$^{\,2}$  and S. Pacheva \footnotemark
\footnotetext{On leave from: Institute of Nuclear Research and Nuclear
Energy, Boul. Tsarigradsko Chausee 72, BG-1784 $\;$Sofia,
Bulgaria. }}
\par \vskip .1in \noindent
Department of Physics, Ben-Gurion University of the Negev \\
Box 653, IL-84105 $\;$Beer Sheva, Israel \\
{\em e-mail}: emil@bguvms.bitnet, svetlana@bgumail.bgu.ac.il
\end{center}
\vskip .2in
\begin{center}
A.H. Zimerman\footnotemark
\footnotetext{Work supported in part by CNPq}
\par \vskip .1in \noindent
Instituto de F\'{\i}sica Te\'{o}rica-UNESP\\
Rua Pamplona 145\\
01405-900 S\~{a}o Paulo, Brazil
\par \vskip .3in

\end{center}

\begin{abstract} \noindent
We show that the 2-matrix string model corresponds to a
coupled system of $2+1$-dimensional KP and modified KP ($\KPm$)
integrable equations subject to a specific ``symmetry'' constraint.
The latter together with the Miura-Konopelchenko map for $\KPm$
are the continuum incarnation of the matrix string equation. The $\KPm$ Miura
and B\"{a}cklund transformations are natural consequences of the underlying
lattice structure. The constrained $\KPm$ system is equivalent to a
$1+1$-dimensional generalized
KP-KdV hierarchy related to graded ${\bf SL(3,1)}$. We provide an explicit
representation of this hierarchy, including the associated
${\bf W(2,1)}$-algebra of the second Hamiltonian structure, in terms of
free currents.
\end{abstract}

\end{titlepage}

\noindent
{\large {\bf 1. Introduction}}
\lskip
One of the remarkable features of integrable Toda lattice hierarchies,
describing the multi-matrix string models \ct{matrix}, is the existence of a
natural elegant way \ct{BX93-94} to extract from them continuum differential
integrable systems without taking any continuum (double-scaling) limit.
These systems are of generalized KP-KdV type and are associated with one
fixed lattice site. This proves to be an efficient approach for
exact calculations of multi-matrix model correlation functions and
exhibits deep relations with topological field theory. In particular,
the two-matrix model in the formalism of \ct{BX93-94} provides a
systematic unified framework for exact treatment via integrable
hierarchies of $c=1$ string theory (for related approaches, cf. \ct{c=1}).
The principal instrument are the $\Win1$ constraints on the two-matrix
model partition function, which are the facets of the pertinent string
equation.

The aim of the present note is to construct a differential integrable system
out of the Toda lattice hierarchy, underlying the two-matrix model, which
directly incorporates in itself the whole information of the string
equation, rather than imposing it as an infinite number of constraints on
the discrete model partition  function. We achieve this by assuming that
at least one of the matrix potentials in the two-matrix model is of
{\em finite} order $p$. In the simplest nontrivial $p=3$ case
the continuum system, we are looking for,
turns out to be a coupled system of $2+1$-dimensional KP and modified KP
($\KPm$) integrable equations subject to two additional constraints.
The latter are
identified as the known Miura-Konopelchenko map \ct{Konop} and a specific
KP ``symmetry'' constraint \ct{constr-KP}. We demonstrate that these two
differential constraints embody the matrix string equation. In particular,
the two-matrix model ``succeptibility'' satisfies the above string-constrained
$\KP3$ equation.

Lattice translations between different sites in the context of Toda hierarchy
are shown to generate the well-known $\KPm$ B\"{a}cklund transformations
\ct{Chau,Oevel-Rogers} which are canonical symmetry of the $\KPm$
Hamiltonian system. Furthermore,
{}from the constrained $2+1$-dimensional (m)KP system we derive an equivalent
{\em unconstrained} $1+1$-dimensional integrable system of generalized KP-KdV
type related to graded ${\bf SL(3,1)}$ algebra \ct{Yu,office}. Systems of
this type were previously obtained in \ct{BX93a,BX93b} by imposing {\sl ad hoc}
second-class constraints on multi-boson KP hierarchies. The corresponding
second Hamiltonian structure is the nonlinear ${\bf W(2,1)}$ algebra,
generalizing the $\nWinf$ algebra \ct{Wu-Yu}. We provide an explicit
free-field representation of ${\bf W(2,1)}$ (cf. also \ct{no2rabin,office}).

\lskip
{\large {\bf 2. Discrete Integrable Systems from Two-Matrix String Models}}
\lskip
This section starts by collecting some basic results on the discrete
integrable systems -- generalized Toda-like lattice systems, which are
associated with multi-matrix string models \ct{BX93-94}. The
associated Toda matrices contain in general {\em finite} number of non-zero
diagonals unlike the case of the full generalized Toda lattice hierarchy
\ct{U-T}.
We present the relevant solutions for the Toda matrices in a form appropriate
for establishing connection with $2+1$-dimensional integrable models.

Specifically, we shall consider the two-matrix model with partition function :
\be
Z_N \lb t,{\ti t},g \rb = \int dM_1 dM_2 \exp -\lcurl
\sum_{r=1}^{p_1} t_r \Tr M_1^r +
\sum_{s=1}^{p_2} {\ti t}_s \Tr M_2^s + g \Tr M_1 M_2 \rcurl   \lab{2-1}
\ee
where $M_{1,2}$ are Hermitian $N \times N$ matrices,
and the orders of the matrix ``potentials'' $p_{1,2}$ may be
finite or infinite. In ref.\ct{BX93-94} it was shown that, by using the method
of generalized orthogonal polynomials \ct{ortho-poly}, the partition function
\rf{2-1} and its derivatives w.r.t. the parameters $\( t_r,{\ti t}_s ,g \)$
can be explicitly expressed in terms of solutions to generalized Toda-like
lattice systems associated with \rf{2-1}. The corresponding linear problem and
Lax (or ``zero-curvature'') representation of the latter read \ct{BX93-94} :
\br
Q_{nm} \psi_m = \l \psi_n  \quad , \quad
-g{\bar Q}_{nm} \psi_m = \partder{}{\l} \psi_n   \lab{L-1} \\
\partder{}{t_r} \psi_n = \( Q^r_{(+)}\)_{nm} \psi_m    \quad , \quad
\partder{}{{\ti t}_s} \psi_n = - \( {\bar Q}^s_{-}\)_{nm} \psi_m   \lab{L-2} \\
\partder{}{t_r} Q = \llb Q^r_{(+)} , Q \rrb  \quad , \quad
\partder{}{{\ti t}_s} Q = \llb Q , {\bar Q}^s_{-} \rrb  \lab{L-3} \\
g \llb Q , {\bar Q} \rrb = \one   \phanta   \lab{string-eq}  \\
\partder{}{t_r} {\bar Q} = \llb Q^r_{(+)} , {\bar Q} \rrb  \quad , \quad
\partder{}{{\ti t}_s} {\bar Q} = \llb {\bar Q} , {\bar Q}^s_{-} \rrb \lab{L-4}
\er
The subscripts $-/+$ denote lower/upper
triangular parts, whereas $(+)/(-)$ denote upper/lower triangular plus
diagonal parts.
The parametrization of the matrices $Q$ and ${\bar Q}$ is as follows:
\br
Q_{nn} = a_0 (n) \quad , \quad Q_{n,n+1} =1 \quad ,\quad
Q_{n,n-k} = a_k (n) \quad k=1,\ldots , p_2 -1   \nonu  \\
Q_{nm} = 0 \quad {\rm for} \;\;\; m-n \geq 2 \;\; ,\;\; n-m \geq p_2 \phanta
\lab{param-1}  \\
{\bar Q}_{nn} = b_0 (n) \quad , \quad {\bar Q}_{n,n-1} = R_n \quad , \quad
{\bar Q}_{n,n+k} = b_k (n) R_{n+1}^{-1} \ldots R_{n+k}^{-1}
\quad k=1,\ldots ,p_1 -1    \nonu  \\
{\bar Q}_{nm} = 0 \quad {\rm for} \;\;\; n-m \geq 2 \;\; ,\;\; m-n \geq p_1
\phanta    \lab{param-2}
\er

{}From now on we shall consider the first evolution parameters $t_1 ,{\ti t}_1$
as coordinates of a two-dimensional space, {\sl i.e.}, ${\ti t}_1 \equiv x$ and
$t_1 \equiv y$, so all modes $a_k (n) , b_k (n)$ and $ R_n $ depend on
$\( x,y ;t_2 ,\ldots ,t_{p_1};{\ti t}_2 ,\ldots ,{\ti t}_{p_2} \)$ .

The second lattice equation of motion \rf{L-4} for $s=1$, using
parametrization \rf{param-2}, gives :
\br
\pa_x R_n &=& R_n \( b_0 (n) - b_0 (n-1) \)\quad,\quad
\pa_x b_0 (n)= b_1 (n) - b_1 (n-1)    \lab{motionx-0} \\
\pa_x \( \frac{b_k (n)}{R_{n+1} \ldots R_{n+k}} \)& =&
\frac{b_{k+1}(n) - b_{k+1}(n-1)}{R_{n+1} \ldots R_{n+k}}  \quad , \;\;
k \geq 2   \lab{motionx-k}
\er
Similarly, the first lattice equation of motion \rf{L-4} for $r=1$ gives :
\be
\pa_y b_0 (n) = R_{n+1} - R_n    \quad,\quad
\pa_y b_k (n) = R_{n+1} b_{k-1}(n+1) - R_{n+k} b_{k-1} (n)
\lab{motiony-k}
\ee
for $k \geq 1$.
{}From the above equations one can express all $b_k (n \pm \ell) \;\; , \;
k \geq 2\,$ and $ R_{n \pm \ell}\,$ ($\ell$ -- arbitrary integer) as
functionals of $b_0 (n) , b_1 (n)$ at a {\em fixed} lattice site $n$.

In complete analogy, the lattice equations of motion \rf{L-3} for
$r=1, s=1$ read explicitly:
\be
\pa_x a_0 (n) = R_{n+1} - R_n \quad,\quad
\pa_x a_k (n) = R_{n-k+1} a_{k-1}(n) - R_n a_{k-1}(n-1) \lab{motionxx-k}
\ee
(with $k \geq 1$) and
\be
\pa_y a_0 (n) = a_1 (n+1) - a_1 (n) \quad,\quad
\pa_y \( \frac{a_k (n)}{R_n \ldots R_{n-k+1}}\) =
\frac{a_{k+1}(n+1) - a_{k+1}(n)}{R_n \ldots R_{n-k+1}}
\lab{motionyy-k}
\ee
with $k \geq 1 $.
Following \rf{motionx-0}, \rf{motiony-k}, \rf{motionxx-k} we obtain the
``duality'' relations:
\be
\pa_y b_1 (n) = \pa_x R_{n+1}
\quad , \quad\pa_x a_0 (n) = \pa_y b_0 (n)
\quad , \quad \pa_x a_1 (n) = \pa_y R_n
\lab{dual}
\ee
{}From the above one gets the two-dimensional Toda lattice equation:
\be
\pa_y \ln R_n = a_0 (n) - a_0 (n-1)
\quad \to \quad
\pa_x \pa_y \ln R_n = R_{n+1}  - 2 R_{n} + R_{n-1}  \lab{motion-00}
\ee

Eqs.\rf{motionxx-k}--\rf{motion-00} allow to express all $a_k (n \pm \ell )
\;\; ,\; k \geq 1$ and $R_{n \pm \ell}$ as functionals of $a_0 (n)$ and
$R_n$ (or $a_1 (n)$ instead) at a fixed lattice site $n$. Furthermore,
due to eqs.\rf{motionxx-k} and \rf{motiony-k} for $k=1$, all
matrix elements of $Q$ and ${\bar Q}$ are functionals of $b_0 (n) , b_1 (n)$
at a fixed lattice site $n$.
Alternatively, due to \rf{dual} we can consider $a_0 (n)$ and $R_{n+1}$ as
independent functions instead of $b_0 (n),b_1 (n)$.

Let us also add the explicit expressions for the flow
eqs. for $b_0 (n), b_1 (n), R_{n+1}$ resulting from \rf{L-3} and \rf{L-4} :
\br
\partder{}{{\ti t}_s}b_0 (n) = \pa_x \( {\bar Q}^s\)_{nn} \quad ,\quad
\partder{}{{\ti t}_s}b_1 (n) =
\pa_x \Bigl\lb R_{n+1}\( {\bar Q}^s\)_{n,n+1}\Bigr\rb    \quad ,\quad
\partder{}{{\ti t}_s}R_{n+1} = \pa_x \( {\bar Q}^s\)_{n+1,n}
\lab{t-s-eqs}\\
\partder{}{t_r}b_0 (n) = \pa_x \( Q^r\)_{nn} \quad ,\quad
\partder{}{t_r}b_1 (n) =
\pa_x \Bigl\lb R_{n+1}\( Q^r\)_{n,n+1}\Bigr\rb \quad ,\quad
\partder{}{t_r}R_{n+1} = \pa_x \( Q^r\)_{n+1,n} \lab{t-r-eqs}
\er
\lskip
{\large {\bf 3. Solution of String Equation for Finite-Order Matrix
Potentials}}
\lskip
{}From now on we assume that one of the matrix potentials in \rf{2-1},
{\sl e.g.}, the second one has a finite order $p_2$, which implies that
the matrix $Q$ has a finite number of diagonals below the main diagonal
(cf. \rf{param-1}). In this case the lattice equations of motion impose
additional constraints relating
the two independent functions $b_0 (n)$ and $b_1 (n)$ (or $R_{n+1}$).

More precisely, we find that for $p_2 < \infty$ the lower triangular plus
diagonal part of $Q$ is expressed through ${\bar Q}$ on the space of
solutions to
\rf{motionxx-k}--\rf{motion-00} in the following form (in what follows, we
shall ignore arbitrary integration constants without loss of generality) :
\be
Q = {\bar Q}^{p_2 -1}_{(-)} + \( {1\over g}x\) \one + I_{+}   \lab{2-3}
\ee
where $g$ is the two-matrix model coupling parameter appearing
in the string equation \rf{string-eq}, and ${I_{+}}_{nm}= \d_{n+1,m}$ .
In particular, for the lowest non-zero diagonal of $Q$ one has:
\be
a_{p_2 -1}(n) = R_n \ldots R_{n -p_2 +2} =
{\bar Q}^{p_2 -1}_{n,n-(p_2 -1)}  \lab{2-4}
\ee
which follows from lattice eq.\rf{motionyy-k} for $k=p_2 -1$ and from the
explicit parametrization of ${\bar Q}$ \rf{param-2}.

Eq.\rf{2-3} can be proved by induction w.r.t. $k=p_2 -1,\ldots,1,0\,$
starting from \rf{2-4}, upon comparing eqs.\rf{motionxx-k}--\rf{motionyy-k}
with the component form of the matrix equations of motion
$\, \pa_y {\bar Q}^s = \sbr{Q_{(+)}}{{\bar Q}^s}\,$ and
$\, \pa_x {\bar Q}^s = \sbr{{\bar Q}^s}{{\bar Q}_{-}}\,$ for arbitrary
integer power $s$.

The remaining lattice equations of motion -- the first equations
of \rf{motionxx-k} and \rf{motion-00},
imply the following two additional constraints on the independent functions
$b_0 (n)$ and $R_{n+1}$ (or $b_1 (n)$)
\be
\pa_y b_0 (n) = {1\over g} + \pa_x \( {\bar Q}^{p_2 -1}_{nn}\) \quad,\quad
\pa_y R_{n+1}= \pa_x \( {\bar Q}^{p_2 -1}_{n+1,n}\)  \lab{constr-0}
\ee

In fact, eqs.\rf{constr-0} are nothing but the component form
of the string eq.\rf{string-eq}. Indeed, it can be rewritten in the form (upon
using \rf{2-3} and \rf{L-3},\rf{L-4}) :
$\( \pa_y - \pa/\pa{\ti t}_{p_2 -1} \) {\bar Q} = {1\over g} \one$ ,
{\sl i.e.}
\be
\partder{}{{\ti t}_{p_2 -1}} b_0 (n) = \pa_y b_0 (n) - {1\over g} \quad ,\quad
\partder{}{{\ti t}_{p_2 -1}} R_{n+1} = \pa_y R_{n+1}  \quad ,\quad
\partder{}{{\ti t}_{p_2 -1}} b_k (n) = \pa_y b_k (n) \lab{string-eq+}
\ee
for $k \geq 1$. Now, inserting \rf{t-s-eqs} into \rf{constr-0} we find that
the latter two equations precisely coincide with the $nn$ and $n+1,n$
component of the string eq.\rf{string-eq+}.
In particular, the evolution parameter ${\ti t}_{p_2 -1}$ is not independent
but rather essentially coincides with $y \equiv t_1$.
It turns out in what follows that it is more convenient to use
${\bar y} \equiv {\ti t}_{p_2 -1}$ as a second space coordinate instead of $y$.

To conclude this section, let us note that there is a complete duality
under $\, p_1 \longleftrightarrow p_2\,$ when the order $p_1$ of the first
matrix potential in \rf{2-1} is also finite. We obtain the exact analogues of
eqs.\rf{2-3}, \rf{constr-0} and \rf{string-eq+} by interchanging
$\, p_1 \longleftrightarrow p_2 \; ,\;
x \equiv {\ti t}_1 \longleftrightarrow t_1 \equiv y \; ,\;
{\ti t}_{p_2 -1} \longleftrightarrow t_{p_1 -1} \; ,\;
Q_{(-)} \longleftrightarrow {\bar Q}_{(+)}$.

\lskip
{\large {\bf 4. Derivation of $\mKP3$ and $\KP3$ from Lattice Systems}}
\lskip
Using the parametrization \rf{param-1}--\rf{param-2}, the equations of the
auxiliary linear Lax problem \rf{L-1},\rf{L-2} acquire the form:
\br
\l \psi_n &=& \psi_{n+1} + a_0 (n) \psi_n + \sum_{k=1}^{p_2 -1}a_k
(n)\psi_{n-k}
\lab{3-1} \\
-{1\over g}\partder{}{\l} \psi_n &=& R_n \psi_{n-1} + b_0 (n) \psi_n +
\sum_{k=1}^{p_1 -1} \frac{b_k (n)}{R_{n+1} \ldots R_{n+k}} \psi_{n+k}
\lab{3-2} \\
\pa_x \psi_n &=& - R_n \psi_{n-1} \quad , \quad
\pa_y \psi_n = \psi_{n+1} + a_0 (n) \psi_n    \lab{3-3}
\er
In particular, one has from the first eq.\rf{3-3} :
\be
\psi_{n-k} = \frac{(-1)^k}{R_{n-k+1}} \pa_x \frac{1}{R_{n-k+2}} \pa_x \ldots
\frac{1}{R_{n}} \pa_x \psi_n  \lab{3-4}
\ee
Inserting \rf{3-3}--\rf{3-4} into \rf{3-1} one obtains the following
{\em two-dimensional} differential Lax spectral equation for $\psi_n$ at a
{\em fixed} lattice site $n$ \foot{In eq.\rf{3-5} we have hidden the spectral
parameter $\l$ by using the identity $\, \partder{}{{\ti t}_{p_2 -1}} \psi_n
\equiv \pa_{\bar y} \psi_n = \( \pa_y - \l \) \psi_n$ , which follows from
\rf{L-1},\rf{L-2} and \rf{2-3}.} :
\br
L_{p_2 -1}(n) \psi_n = 0 \phantc    \lab{3-5} \\
L_{p_2 -1}(n) = \pa_{\bar y} + \sum_{k=1}^{p_2 -1}
\frac{(-1)^k a_k (n)}{R_n \ldots R_{n-k+1}}
\(\pa_x - \pa_x \ln\(R_n \ldots R_{n-k+2}\)\) \cdots
\(\pa_x - \pa_x \ln R_n\) \pa_x \nonu \\
= \pa_{\bar y} + (-1)^{p_2 -1} \pa_x^{p_2 -1} + \sum_{k=1}^{p_2 -2} (-1)^k
f^{(p_2 -1)}_k \Bigl( b_0 (n),b_1 (n) \Bigr) \pa_x^k
\equiv \pa_y  + M_{p_2 -1}(n)    \lab{3-6}
\er
where in the second line of eq.\rf{3-6} we have used the lattice
equations of motion \rf{2-3}, {\sl i.e.},
$Q_{-} = \( {\bar Q}^{p_2 -1}\)_{-}$,
to express the coefficient functions of the 2-dimensional Lax differential
operator in terms of the independent $b_{0,1}(n)$.

Similarly, the lattice flow equations w.r.t. ${\ti t}_s$ \rf{L-2} can be
equivalently written as differential flow equations:
\be
\partder{}{{\ti t}_s} \psi_n = - M_s (n) \psi_n
\quad ,\;\;
M_s (n) \equiv (-1)^s \pa_x^s + \sum_{k=1}^{s-1} (-1)^k
f^{(s)}_k \Bigl( b_0 (n),b_1 (n)\Bigr) \pa_x^k  \lab{3-8-a}
\ee
with $s=2,\ldots ,p_2  $. Explicitly we have for the coefficient functions:
\be
f^{(s)}_{s-1} = s b_0 (n)  \quad , \quad
f^{(s)}_{s-2} = s b_1 (n) +{{s}\choose{2}} \( b_0^2 (n) - \pa_x b_0 (n) \)
\quad , \;\; {\rm etc.}  \lab{3-9}
\ee
In particular, comparing \rf{3-8-a}--\rf{3-9} and \rf{3-5}--\rf{3-6},
we note that the Lax spectral equation \rf{3-5} is in fact the
$(p_2 -1)$-th flow equation \rf{3-8-a}.

The compatibility conditions for \rf{3-8-a} :
\be
\partder{}{{\ti t}_r} M_s (n) - \partder{}{{\ti t}_s} M_r (n)  +
\Bigl\lb M_r (n) ,\, M_s (n) \Bigr\rb = 0  \quad ,
\;\; r,s=2,\ldots p_2 \lab{3-10}
\ee
yield for $r=p_2 -1\,$ ({\sl i.e.}, ${\ti t}_{p_2 -1}={\bar y}\,$)
a system of $2+1$-dimensional integrable nonlinear
evolution equations for $b_0 (n), b_1 (n)$ subject to the additional
string-equation constraints \rf{constr-0}.
Let us consider in more detail the simplest nontrivial case $p_2 =3$ where:
\be
M_2 (n) =  \pa_x^2 - 2b_0 (n) \pa_x  \;\; , \;\;
M_3 (n) =  - \pa_x^3 + 3 b_0 (n) \pa_x^2
- 3 \( b_1 (n) + b_0^2 (n) - \pa_x b_0 (n) \) \pa_x     \lab{3-11-a}
\ee
Eq.\rf{3-10} with $r=2,s=3$ yields firstly the non-dynamical equation:
\be
\pa_{\bar y} b_0 (n) = \pa_x \( 2b_1 (n) - \pa_x b_0 (n) + b_0^2 (n)\)
\lab{Konop}
\ee
which coincides with the first string eq.\rf{constr-0} (for $p_2 = 3$).
Secondly, it yields an evolution equation for $b_0 (n)$ :
\br
4\partder{}{{\ti t}_3} b_0 (n) &=& \pa_x^3 b_0 (n) - 2 \pa_x b_0^3 (n) +
3\pa_{\bar y} {\bar a}_0 (n) + 6 {\bar a}_0 (n) \pa_x b_0 (n)  \lab{mKP3}  \\
\pa_x {\bar a}_0 (n) &=& \pa_{\bar y} b_0 (n) \quad , \quad
{\bar a}_0 (n)\equiv a_0 (n) - {1\over g}x =
2b_1 (n) - \pa_x b_0 (n) + b_0^2 (n)  \lab{3-12}
\er
where we have used \rf{dual} and \rf{Konop}.
Combining \rf{Konop} with \rf{mKP3} yields an evolution equation for
$b_1 (n)$ :
\be
4\partder{}{{\ti t}_3} b_1 (n) = \pa_x^3 b_1 (n) + 6 \pa_x b_1^2 (n) +
3 \pa_{\bar y} R_{n+1} \qquad , \quad \pa_x R_{n+1} = \pa_{\bar y} b_1 (n)
\lab{KP3}
\ee
Finally, the second string eq.\rf{constr-0} (for $p_2 = 3$) yields:
\be
\pa_{\bar y} R_{n+1} = \pa_x \llb \pa_x R_{n+1} + 2 b_0 (n) R_{n+1} \rrb
\equiv M^{\ast}_2 (n) R_{n+1}   \lab{constr-KP}
\ee
where $M^{\ast}_2 (n)$ denotes the operator adjoint to $M_2 (n)$ in
\rf{3-11-a}.

Now, it is straightforward to recognize eq.\rf{KP3}
as the $2+1$-dimensional KP equation $(\KP3 )$, eq.\rf{mKP3} -- as the
modified KP equation $(\mKP3 )$, and eq.\rf{Konop} -- as the associated
$2+1$-dimensional Miura-Konopelchenko map \ct{Konop} relating them.
However, the present two-matrix model derivation reveals
an {\em additional} constraint \rf{constr-KP} relating $\KP3$
and $\mKP3$ equations. This constraint
can be identified as a particular example of a ``symmetry constraint''
\ct{constr-KP} within the
framework of the standard $1+1$-dimensional KP hierarchy $\(\KPt\)$ formalism.
The KP reduction, triggered by \rf{constr-KP}, is studied in section 6.

As demonstrated above, both the $\KPm$ eqs. \rf{mKP3} and
\rf{KP3}, as well as the Miura-Konopelchenko map \rf{Konop} and the
``symmetry constraint'' \rf{constr-KP} KP reduction
naturally arise from the underlying Toda-like lattice integrable structure
associated with the two-matrix string model. Let us emphasize that
the Miura-Konopelchenko map \rf{Konop} and the ``symmetry constraint''
\rf{constr-KP} explicitly embody the ``string equation'' \rf{string-eq}
of the two-matrix model in the
case $p_2 =3$. Thus, they are the $2+1$-dimensional analog of the
$\Win1$-constraints on the partition function \rf{2-1} in the formalism of
refs.\ct{BX93-94}.

Through the method of orthogonal polynomials the two-matrix model partition
function \rf{2-1} is given by:
\be
Z_N \lb t,{\ti t},g \rb  = const \, N! \prod_{n=0}^{N-1} h_n \quad , \quad
h_n = \exp \int^x b_0 (n)  \lab{Z}
\ee
Thus, its calculation reduces to finding the appropriate solutions of
$\KPm$ equations  \rf{mKP3},\rf{KP3} subject to the ``string equation''
constraint \rf{constr-KP}. In particular, as follows from \rf{Z} and
\rf{motionx-0}, the ``succeptibility''
$\pa_x^2 \ln Z_N = b_1 (N-1)$ satisfies the string-constrained
$\KP3$ equation \rf{KP3},\rf{constr-KP}.

\lskip
{\large {\bf 5. Symmetries of Constrained $\KP3$ and $\mKP3$}}
\lskip
In the previous section we represented the differential $\KP3$ and $\mKP3$
integrable systems explicitly in terms of objects from the underlying
generalized Toda-like lattice structure. Now we are going to show that this
underlying lattice structure naturally exhibits the discrete B\"{a}cklund
transformations for $\KP3$ and $\mKP3$.

On the lattice itself we have two obvious discrete symmetry operations:

(a) lattice site translation $ n \lra n-1\,$, in particular, for
the discrete Lax ``wave'' function
\be
\psi_{n+1} \lra \psi_n = -  R_{n+1}^{-1} \pa_x \psi_{n+1}   \lab{symm-a}
\ee

(b) similarity (phase) transformations on the same lattice site, {\sl e.g.},
gauge transformations
\be
\psi_n \longrightarrow {\wti \psi}_n = e^{\phi_n} \psi_n \equiv \Phi_n \psi_n
\lab{symm-b}
\ee

First, we consider type (a) lattice symmetry \rf{symm-a}.
Going back to $\mKP3$ \rf{mKP3} and $\KP3$ \rf{KP3}, we observe that
$b_0 (n)$ and $b_1 (n) $ do satisfy them {\em for any} fixed
site $n$ on the underlying lattice, {\sl i.e.,} if
$b_0 (n), b_1 (n)$ satisfy \rf{mKP3},\rf{KP3}, so do
$b_0 (n-1), b_1 (n-1)$ . On the other hand, from the
lattice equations of motion \rf{motionx-0}, \rf{motiony-k}
we obtain the following simple relations for the lattice shifts
of the above functions:
\br
b^{(-)}_0 (n) \!\!&\equiv& \!\!
b_0 (n-1) = b_0 (n) - \pa_x \ln \( \pa_x \Phi_n\)
\;\;, \;\; b^{(-)}_1 (n) \equiv b_1 (n-1) = b_1 (n) - \pa_x b_0 (n)
\lab{Backlund--b}  \\
\pa_x \Phi_n &=& R_n = R_{n+1} - \pa_{\bar y} b_0 (n) - {1\over g}
\lab{constr-Phi}
\er
which can be viewed as discrete symmetry (B\"{a}cklund) transformations for
$\KPm$ \rf{mKP3} and \rf{KP3}.
Obviously, type (a) B\"{a}cklund transformation
\rf{Backlund--b} preserves the $\KPm$ ``symmetry''
constraint \rf{constr-KP}. Note that the two-matrix model coupling constant
$g$ emerges here as a free B\"{a}cklund parameter.

Let us also note, that the B\"{a}cklund transformation for $b_1 (n)$
\rf{Backlund--b}) (taking into account \rf{Konop}) is a canonical
transformation for the
$2+1$-dimensional KP Hamiltonian system (cf. \ct{Dickey}), {\sl i.e.},
it leaves invariant the canonical Poisson brackets and the Hamiltonian
(here there is no difference between $y$ and ${\bar y}$) :
\br
\pbbr{b_1 (n) (x,y)}{b_1 (n) (x^{\pr},y^{\pr})} =
{1\over {16}}\pa_x \d (x-x^{\pr}) \d (y-y^{\pr}) =
\pbbr{b_1 (n+1) (x,y)}{b_1 (n+1) (x^{\pr},y^{\pr})}   \lab{KP3-PB} \\
H(n) = 2 \int dx\,dy\, \llb \(\pa_x b_1 (n)\)^2 - 4 b_1^3 (n) -
3 R_{n+1}^2 \rrb = H(n+1) \quad ,\;\; \pa_x R_{n+1} = \pa_y b_1 (n) \quad
\lab{KP3-Ham}
\er

In the framework of the standard $\KPt$ hierarchy formalism with $\KPt$
Lax operators:
\be
L_{mKP} = \pa_x + v + \sum_{k \geq 1} v_k \pa_x^{-k}  \quad , \quad
L_{KP} = \pa_x + \sum_{k \geq 1} u_k \pa_x^{-k}     \lab{KP2}
\ee
eqs.\rf{Backlund--b} have already been obtained \ct{Chau,Oevel-Rogers}
(identifying $v = b_0 (n) ,\; u_1 \equiv 2u = 2 b_1 (n)$ ) via the method of
pseudo-differential operator gauge transformations and are known as the first
type of auto-B\"{a}cklund transformations for conventional $\KPm$. However,
in the present two-matrix model realization of $\mKP3$ and $\KP3$
there is
an additional important property of this auto-B\"{a}cklund transformation ---
the additional constraint \rf{constr-Phi} expressing $\Phi \equiv \Phi_n$ as a
specific explicit functional of the $\KPm$ functions $u,v$ .
This additional constraint on the
auto-B\"{a}cklund transformation results from the $\KPm$ ``string-equation''
constraint \rf{constr-KP}. Both constraints \rf{constr-KP} and \rf{constr-Phi}
are absent in the general $\KPm$ setting.

Next, we consider type (b) lattice symmetry \rf{symm-b}.
It must preserve the form of the
2-dimensional Lax and the corresponding ${\ti t}_s$-flow generating operators:
\br
\Phi_n^{-1} \( \partder{}{{\ti t}_s} + M_s (n) \) \(\Phi_n \cdot\) =
\partder{}{{\ti t}_s} + {\wti M}_s (n) \quad ,\;\; s=2,\dots p_2
\phanta \lab{Phi-gauge}  \\
{\wti M}_s (n) \equiv (-1)^s \pa_x^s + \sum_{k=1}^{s-1} (-1)^k
f^{(s)}_k \( {\wti b}_0, {\wti b}_1 \) \pa_x^k  \;\;\; ,
\;\; {\wti b}_0 (n) = b_0 (n) - \pa_x \ln \Phi_n  \;\; , \;\;
{\wti b}_1 (n) = b_1 (n)  \lab{Backlund2}
\er
where the last two equations represent type (b) B\"{a}cklund transformation
for unconstrained $\KPm$ \ct{Chau,Oevel-Rogers}.
Comparing \rf{Phi-gauge} with the similarity relation between the operators
$M_s (n)$ and $M_s (n+1)$ implied by type (a) lattice symmetry \rf{symm-a} :
\be
\frac{1}{R_{n+1}} \pa_x \llb \partder{}{{\ti t}_s} + M_s (n+1) \rrb
\pa_x^{-1} \( R_{n+1} \cdot \) = \partder{}{{\ti t}_s} + M_s (n)
\quad ,\;\; s=2,\dots p_2   \lab{similar}
\ee
we find that $\Phi_n$ in \rf{Phi-gauge}--\rf{Backlund2} is again given by
eq.\rf{constr-Phi}. However, with such $\Phi_n$
type (b) B\"{a}cklund transformation \rf{Backlund2}
{\em does not} preserve the string-equation constraint \rf{constr-KP}.
Therefore, only type (a) (lattice translation) B\"{a}cklund transformation
\rf{Backlund--b} survives in the two-matrix-model realization of $\KPm$.

Concluding this section, let us consider also the continuum symmetries of the
constrained $\KPm$. In this respect we
note further important differences between the string matrix-model
realization of $\mKP3$ and $\KP3$ integrable systems, on one hand,
and the standard $\KPt$ integrable hierarchy, on the other hand.
In the present $2+1$-dimensional matrix-model formalism the number $p_2$ of
the usual KP time-evolution parameters (${\ti t}_1 = x,{\ti t}_2 ,\ldots ,
{\ti t}_{p_2 -1}={\bar y}, {\ti t}_{p_2}$) is {\em finite},
{\sl e.g.}, $p_2 = 3$ in the above analysis, unlike the $\KPt$ case.
Furthermore, besides them there is also an additional set of time-evolution
parameters $t_1\equiv y, t_2 ,\ldots ,t_{p_1}$, which might be {\em infinite}
in number, {\sl i.e.}, $p_1 \lra \infty\,$. Their corresponding flows,
given by \rf{t-r-eqs}, are both commuting with the ${\ti t}_s$-flows
(given by \rf{t-s-eqs}), as well as commuting among
themselves. So the latter correspond to an (infinite-dimensional) Abelian
symmetry algebra for the constrained string-matrix-model's $\KPm$ system.

The nature of the continuum symmetries corresponding to the $t_r$-flows
could be better understood if one considers the constrained $\KPm$ in terms
of the equivalent $1+1$ dimensional integrable system
(eqs.\rf{1+1-a}--\rf{1+1-cc} below).
\lskip
{\large {\bf 6. Equivalent ${\bf 1+1}$-Dimensional Formulation of
Constrained \\
$\KPm$ }}
\lskip
The string-constrained $\KPm$ system \rf{mKP3}--\rf{constr-KP}, which
describes the two-matrix model (for $p_2 =3$), can be explicitly reduced to
an equivalent $1+1$-dimensional generalized KP-KdV integrable system.
Upon excluding in \rf{mKP3} ${\bar a}_0 (n)$ via \rf{3-12} and \rf{Konop}, and
substituting \rf{constr-KP} into \rf{KP3},
the corresponding reduced $1+1$-dimensional integrable system can be written
in the following form:
\br
\partder{}{{\ti t}_3} R_{n+1} &=& \pa_x \Bigl\lb \pa_x^2 R_{n+1} + 3\( b_1 (n)
+
b_0^2 (n) \) R_{n+1} + 3 b_0 (n) \pa_x R_{n+1} \Bigr\rb    \lab{1+1-a} \\
\partder{}{{\ti t}_3} b_0 (n) &=& \pa_x \Bigl\lb \pa_x^2 b_0 (n) + b_0^3 (n) +
3 b_1 (n) b_0 (n) + {3\over 2} R_{n+1}
- {3\over 2} \pa_x \( b_1 (n) + b^2_0 (n) \) \Bigr\rb    \lab{1+1-b} \\
4\partder{}{{\ti t}_3} b_1 (n) &=& \pa_x \Bigl\lb \pa_x^2 b_1 (n) +
6 b_1^2 (n) + 3 \pa_x R_{n+1} + 6 b_0 (n) R_{n+1} \Bigr\rb      \lab{1+1-c}
\er
Similarly, the ${\ti t}_2$ flow eqs. acquire the form:
\br
\partder{}{{\ti t}_2} R_{n+1} &=& \pa_x \Bigl\lb \pa_x R_{n+1} +
2b_0 (n) R_{n+1} \Bigr\rb   \phanta  \lab{1+1-aa} \\
\partder{}{{\ti t}_2} b_0 (n) &=& \pa_x \Bigl\lb 2 b_1 (n) + b_0^2 (n) -
\pa_x b_0 (n) \Bigr\rb   \quad ,\quad
\partder{}{{\ti t}_2} b_1 (n) =  \pa_x R_{n+1}   \lab{1+1-cc}
\er

In complete analogy with the $2+1$-dimensional type (a) B\"{a}cklund
transformation \rf{Backlund--b}, from the lattice equations of
motion one immediately obtains the B\"{a}cklund transformation for the
$1+1$-dimensional system \rf{1+1-a}--\rf{1+1-cc} resulting from negative
lattice site translation:
\br
b^{(-)}_0 (n) &\equiv& b_0 (n-1) =
b_0 (n) - \pa_x \ln \Bigl( R_{n+1} - {1\over g} - \pa_x \( 2b_1 (n) + b^2_0 (n)
- \pa_x b_0 (n) \)\Bigr)    \lab{Backl-1+1-a}  \\
b^{(-)}_1 (n) &\equiv& b_1 (n-1) = b_1 (n) - \pa_x b_0 (n) \lab{Backl-1+1-b}
\\
R^{(-)}_{n+1} &\equiv& R_n =
R_{n+1} - {1\over g} - \pa_x \( 2b_1 (n) + b^2_0 (n) - \pa_x b_0 (n) \)
\lab{Backl-1+1-c}
\er
Again, as in \rf{Backlund--b}, the two-matrix model
coupling constant $g$ appears as a free B\"{a}cklund parameter.

The system \rf{1+1-a}--\rf{1+1-c} (or \rf{1+1-aa}--\rf{1+1-cc}) possesses Lax
representation which is $1+1$-dimensional analog of
\rf{3-5}--\rf{3-6} (with $p_2 =3$) :
\br
\l \psi_n &=& {\wti L}_{p_2 -1}(n) \psi_n  \quad , \quad
\partder{}{{\ti t}_s} \psi_n = - M_s (n) \psi_n
\quad ,\;\; s=2,\ldots ,p_2   \phanta \lab{5-1}    \\
{\wti L}_{p_2 -1}(n) &=& - \pa_x^{-1} \( R_{n+1} \cdot \) +
{\bar Q}^{p_2-1}_{nn} + M_{p_2 -1}(n) \phanta  \lab{5-2} \\
{\wti L}_{2}(n) &=& - \pa_x^{-1} \( R_{n+1} \cdot \) + 2 b_1 (n) +
\( \pa_x - b_0 (n) \)^2  \quad ,\;\; {\rm for} \; p_2 =3 \phanta \lab{5-3}
\er
In particular, one reproduces the lattice flow
eqs.\rf{t-s-eqs} as compatibility conditions for the
linear auxiliary Lax problem \rf{5-1}.
The systems \rf{1+1-a}--\rf{1+1-c} and \rf{1+1-aa}--\rf{1+1-cc} themselves
arise as compatibility conditions:
\be
\partder{}{{\ti t}_3} {\wti L}_{2}(n) + \llb M_3 (n)\, ,\, {\wti L}_{2}(n)\rrb
= 0 \quad , \quad
\partder{}{{\ti t}_2} {\wti L}_{2}(n) + \llb M_2 (n)\, ,\, {\wti L}_{2}(n)\rrb
= 0  \lab{Lax-1+1}
\ee
with ${\wti L}_{2}(n)$ as in \rf{5-3}, and $M_{2,3} (n)$ as in
\rf{3-11-a}.

Let us concentrate on the Lax operator \rf{5-3} \foot{To avoid confusion,
{}from here on $D$ will denote the differential operator w.r.t. $x$, whereas
$\pa_x f$ will denote derivative acting on a function. Note also the inverse
ordering of differential operators and coefficient functions in
the definiton of the Lax operators.}
(indication of the associated lattice site $n$ is suppressed for brevity) :
${\wti L}_2 = -D^{-1}\, R + 2b_1 + b_0^2 + \pa_x b_0 - D \, \(2 b_0\) + D^2$.
The Hamiltonian structures inherent in the integrable system
\rf{1+1-a}--\rf{1+1-c} (or \rf{1+1-aa}--\rf{1+1-cc}) can naturally be
identified using the general $R$-matrix scheme of
Adler-Kostant-Symes/Reyman-Semenov-Tian-Shansky
\ct{AKS-RST,MST83} in the context of the algebra of all pseudo-differential
operators $\PsDA$ . Namely, the first and the second Hamiltonian
structures are given in terms of the following $R$-matrix Poisson brackets:
\br
\pbr{\me{L}{X}}{\me{L}{Y}}_R^{(1)}
&= &- \me{L}{\h \sbr{RX}{Y} + \h \sbr{X}{RY}}  \lab{1st-Rbra}  \\
\pbr{\me{L}{X}}{\me{L}{Y}}_R^{(2)} &=& - \me{L}{\h \sbr{R(XL)}{Y} +
\h \sbr{X}{R(LY)}}     \lab{2nd-Rbra}
\er
where $L,X,Y$ are arbitrary elements of $\PsDA$:
$L = \sum_{k \geq -\infty} D^k\, u_k$ and
$X = \sum_{k \geq - \infty} X_k D^k  $
and $<\cdot \v \cdot >$ denotes
the standard bilinear pairing via the Adler trace.
$\me{L}{X} = {\Tr}_A \( LX\)  = \Intres \( LX\)$.
Both Hamiltonian structures are compatible provided $R$ is antisymmetric
operator on $\PsDA$ w.r.t. the Adler trace.
There are three
different natural $R$-matrix operators on $\PsDA\,$ labelled by
$\ell = 0,1,2$ \ct{Reyman} :
\be
R_{\ell} = P_{\geq {\ell} } - P_{\leq {\ell} - 1} \quad,\quad
P_{\geq \ell} X = \sum_{k\geq \ell} X_k D^k \quad , \quad
P_{\leq \ell -1} X = \sum_{k= \ell -1}^{\infty} X_{-k} D^{-k}
\lab{R-mat}
\ee

Now, one observes (cf. \ct{Kuper85,ANPV})
that the space of Lax operators ${\wti L}_2 $
($p_2 =3$),
and more generally -- the Lax operators for arbitrary (finite) $p_2$
\rf{5-2} :
\be
{\wti L}_{p_2 -1} = -D^{-1}\, R + \sum_{k=0}^{p_2 -2} (-1)^k D^k \,
{\wti f}_k^{(p_2 -1)} + (-1)^{p_2 -1} D^{p_2 -1}
\lab{Laxop-all}
\ee
with ${\wti f}_k^{(p_2 -1)}$ simply expressed in terms of
$f_k^{(p_2 -1)}$ from \rf{3-9} for $s=p_2 -1$,
span a {\em closed} $R_1$-coadjoint orbit on $\PsDA$ with $R_1$ as in
\rf{R-mat}.
The corresponding first Hamiltonian structure is given by the
$R$-Kirillov-Kostant bracket \rf{1st-Rbra} with $L= {\wti L}_2$ and
$R=R_1=P_{\geq 1} - P_{\leq 0}$.
The ${\ti t}_3$- and ${\ti t}_2$-flow eqs.\rf{1+1-a}--\rf{1+1-c} and
\rf{1+1-aa}--\rf{1+1-cc} are Hamiltonian equations of motion w.r.t. the
first bracket \rf{1st-Rbra}, which reads explicitly
$\pbr{R(x)}{b_0 (x^{\pr})} = - \pa_x \d (x-x^{\pr}) $
and
$\pbbr{b_1 (x)}{b_1 (x^{\pr})} = - \h \pa_x \d (x-x^{\pr})$, and
with Hamiltonians\foot{The overall
minus sign in the definition of $H_N$
is due to the inverse ordering of $D$'s and
coefficient functions in ${\wti L}_2 $ .}
$\, H_N = -{1\over N} {\Tr}_A {\wti L}_2^N \,$ for $N={5\over 2},2$ .

Concerning the construction of the second Hamiltonian structure, we note
that $R_1$ from \rf{R-mat} is {\em not} antisymmetric.
However, there exists a {\em symplectic} mapping
\ct{Kuper85,ANPV,Oevel-Rogers}
{}from the $R_1$-brackets to the standard $R_0$-brackets of the first KP
Hamiltonian structure where now $R_0$ \rf{R-mat} is already antisymmetric:
\be
\Pbr{\bil{{\wti L}_2}{X}}{\bil{{\wti L}_2}{Y}}_{R_1}^{(1)} =
\Pbr{\bil{{\wti L}_2^{(0)}}{X}}{\bil{{\wti L}_2^{(0)}}{Y}}_{R_0}^{(1)}
\quad,\quad
{\wti L}_2^{(0)} = e^{- \int_x b_0} {\wti L}_2 e^{\int_x b_0}  \lab{Laxop-0}
\ee
with $ {\wti L}_2^{(0)} =- \( D + b_0 \)^{-1}R + 2b_1 + D^2$.
Therefore, we can now use the general formula \rf{2nd-Rbra} to construct the
second Poisson brackets compatible with the first bracket, however,
with a caution. Namely, let us emphasize the following important point.
Eq.\rf{2nd-Rbra} is given for {\em generic} pseudo-differential Lax operators
$L$. Restriction of $L$ to arbitrary submanifold {\em does not}
necessarily lead to a consistent restriction of the corresponding $R$-matrix
Poisson brackets \rf{2nd-Rbra} (see, {\sl e.g.} \ct{Morosi}).
Thus, we have to prove that the restriction of generic $L$ to
${\wti L}_2^{(0)}$ \rf{Laxop-0} is in fact a
consistent Poisson reduction. This can be done by adapting the proof in
\ct{no2rabin} for the multi-boson reductions of the ordinary KP hierarchy.
The final result for the second Hamiltonian structure for ${\wti L}_2^{(0)}$
reads \foot{The last term on the r.h.s. of \rf{2nd-bra} is a standard
Dirac-bracket term due to the absence of next-to-leading differential order
term in ${\wti L}_2^{(0)}$ \rf{Laxop-0}.} :
\br
\Pbr{\bil{{\wti L}_2^{(0)}}{X}}{\bil{{\wti L}_2^{(0)}}{Y}}^{(2)} =
{\Tr}_A \Bigl( \( {\wti L}_2^{(0)}X\)_{\geq 0} {\wti L}_2^{(0)}Y -
\( X {\wti L}_2^{(0)}\)_{\geq 0} Y {\wti L}_2^{(0)} \Bigr)  \nonu \\
+ \h \int dx \, {\rm Res} \Bigl( \sbr{{\wti L}_2^{(0)}}{X}\Bigr) \pa_x^{-1}
{\rm Res} \Bigl( \sbr{{\wti L}_2^{(0)}}{Y}\Bigr)       \lab{2nd-bra}
\er
and the specific expressions for the nonlinear Poisson brackets among the
coefficient fields can be found in refs.\ct{BX93a,BX93b,office}.

It follows from the general considerations in \ct{BX93b}, that the
coefficient fields $R, b_0 , b_1$ generate the ${\bf W(2,1)}$-algebra, which is
a generalization of the nonlinear $\nWinf$-algebra \ct{Wu-Yu}.
On the other hand, in \ct{no2rabin} we succeeded to represent the
usual multi-boson KP hierarchies in terms of canonical pairs of free fields
abelianizing the second Hamiltonian structure of ordinary KP hierarchy.
Using the latter result, we find a similar result for ${\wti L}_2^{(0)}$ --
explicit expressions of the coefficient fields $\( R_{n+1},b_0 (n),b_1 (n)\)$
in terms of free fields (currents) $\( c_1 ,e_1 , c_2 \)$ :
\br
R = \( \pa_x + e_1 + c_1 + 2 c_2 \) \( \pa_x + c_1 \) e_1  \phanta \nonu \\
b_1 = \h \( \pa_x + c_1 \) e_1 - \h \( \pa_x + c_2 \) c_2 \quad , \quad
b_0 = e_1 + c_1 + c_2  \qquad\quad  \lab{free-field}  \\
\pbbr{c_1 (x)}{e_1 (x^\pr )} = - \pa_x \d \( x-x^\pr \) \quad , \quad
\pbbr{c_2 (x)}{c_2 (x^\pr )} = \h \pa_x \d \( x-x^\pr \) \quad , \quad
rest = 0  \lab{free-pb}
\er
In turn, formulas \rf{free-field},\rf{free-pb} provide an explicit free-field
representation of ${\bf W(2,1)}$-algebra embodied in the
second bracket structure
\rf{2nd-bra}. Let us note, that \rf{free-field}--\rf{free-pb} can also be
obtained from the free current's realizations of the modified full KP
hierarchy \ct{Yu} by a specific Dirac constraint reduction. Similar
free-field representations exist for ${\bf W(p_2 -1,1)}$ Poisson-bracket
algebras for the higher Lax operators \rf{Laxop-all} which describe
generalized graded ${\bf SL(p_2 ,1)}$ KP-KdV integrable systems
\ct{Yu,office}.

We conclude with a remark about the infinite-dimensional Abelian symmetry
algebra of the $1+1$-dimensional integrable system
\rf{1+1-a}--\rf{1+1-cc}, \rf{5-3}--\rf{Lax-1+1}, generated by the $t_r$-flows
\rf{t-r-eqs}. Inserting the solution for $Q$ \rf{2-3} into \rf{t-r-eqs}, we
find that the corresponding vector fields $\partder{}{t_r}$
are Hamiltonian ones, with Hamiltonian functions $\cH_r$ w.r.t. the first
Poisson bracket \rf{1st-Rbra} of the following form: $\cH_r = {\bar H}_{r+1}=
- {1\over {r+1}} {\Tr}_A {\bar L}_2^{r+1}\,$ for integer $r \geq 1$. The
Lax operator ${\bar L}_2$ is of the same form as \rf{5-2}--\rf{5-3}, but
with $a_0 (n) = {\bar Q}^2_{nn} + {1\over g}x$ as a zero-order term instead of
${\bar Q}^2_{nn} \equiv 2 b_1 (n) + b_0^2 (n) - \pa_x b_0 (n)$ .

\lskip
\underbar{Outlook} ~
It would be interesting to extend the analysis of sections 4--6 beyond
the simplest case of $p_2=3$ and for higher multi-matrix models.
The general case of Lax operators ${\wti L}_{p_2 -1}$
\rf{Laxop-all} with arbitrary finite differential part of order $p_2 -1$
is studied in ref.\ct{office}.
{}From eq.\rf{Laxop-all} we expect then to find explicitly the generalized
$SL(p_2,q)$-KdV hierarchies with $q$ restricted to $q=1$.
One recalls at this point that the generalized Kontsevich models
\ct{Kont} provide description of the continuous models
(based on $c<1$ minimal conformal models) only for $q=1$ in the
$(p,q)$-series and it would therefore be natural to study a
possible relation.

\lskip
{\bf Acknowledgements.}
E.N. and S.P. acknowledge illuminating discussions with V. Ogievetsky.
A.H.Z. thanks UIC for hospitality and FAPESP for financial support.

\end{document}